\documentclass[3p,times]{elsarticle}

\usepackage{ecrc}

\volume{00}

\firstpage{1}

\journalname{Nuclear Physics A}

\runauth{S. Masciocchi et al.}

\jid{nupha}

\usepackage{amssymb}

\usepackage[figuresright]{rotating}
\usepackage{graphicx}

\providecommand*\pp{pp~}
\providecommand{\s}{$\sqrt{s}$~}

\providecommand{\pt}{\ensuremath{p_{\rm t}}~}
\providecommand*\gevc{GeV/$c$~}

\begin{document}

\begin{frontmatter}




\title{Heavy-flavour production in ALICE at the LHC}


\author{S. Masciocchi for the ALICE Collaboration}
\ead{s.masciocchi@gsi.de}
\address{EMMI and GSI Helmholtzzentrum fuer Schwerionenforschung GmbH,
Darmstadt, Germany}

\begin{abstract}
ALICE at the LHC is the experiment dedicated to study the physics of nucleus-nucleus collisions. 
The apparatus is well suited for
the measurement of heavy-quark hadron production, making use of the high
spatial resolution provided by the tracking detectors and the excellent particle identification, which
are distinctive of the ALICE apparatus.

Results from proton-proton collisions at \s = 2.76 and 7 TeV, and from Pb--Pb collisions at 
{\mbox {$\sqrt{s_{\rm NN}}$ = 2.76 TeV}} are presented. The measurements in pp collisions 
provide an important test of perturbative QCD predictions. The precise vertex reconstruction together 
with the electron identification, allows the separation of the charm and the beauty components.
Furthermore, the pp results are essential as a reference for the measurements in heavy-ion collisions. 
Nuclear modification factors were measured for D mesons, for electrons and for muons from heavy-flavour hadron
decays. The elliptic flow of D mesons is also discussed. 
These measurements are important because they will 
provide information on the Quark-Gluon Plasma
produced in heavy-ion collisions, 
via the energy loss of the heavy partons in the strongly interacting medium, and hints on the medium thermalization. 
\end{abstract}


\end{frontmatter}


\section{Introduction}
\label{intro}

Heavy quarks, charm and beauty, are excellent probes to investigate quantum chromodynamics (QCD)
processes in hadronic interactions, and to characterize the deconfined medium produced 
in high-energy heavy-ion collisions, the Quark-Gluon Plasma (QGP). Because of the large quark masses, heavy-flavour
production proceeds mainly through initial hard parton-parton scattering. In proton-proton collisions,
therefore, the production cross sections of charm and beauty quarks provide a test of perturbative QCD 
(pQCD). They also provide a crucial baseline for corresponding measurements in heavy-ion collisions.
In such collisions a strongly interacting medium is formed, in which heavy quarks interact, after having 
been produced in the very initial stage of the collision. Consequently, heavy quarks experience the whole
history of the medium evolution: they lose energy while propagating through the medium, and they 
might participate in the collective dynamics. A measurement of the azimuthal anisotropy of the charm
hadron production and their elliptic flow ($v_2$) will give information on the charm thermalization
in the medium.

A sensitive observable to characterize the effect of the dense medium on the heavy-quark production is
the nuclear modification factor
$R_{\rm AA}$. This is defined as:

\begin{equation}
R_{\rm AA}(p_{\rm t}) = \frac{1}{\langle T_{\rm AA}\rangle} \times \frac{{\rm d}N_{\rm AA}/{\rm d}p_{\rm t}}{{\rm d}\sigma_{\rm pp}/{\rm d}p_{\rm t}} 
\label{eq:raa}
\end{equation}

\hspace*{-5.2mm}where ${\rm d}N_{\rm AA}/{\rm d}p_{\rm t}$ is the transverse momentum spectrum measured in heavy-ion (A--A) collisions,
${\rm d}\sigma_{\rm pp}/{\rm d}p_{\rm t}$ is the $p_{\rm t}$-differential cross-section measured in pp collisions, and
the ratio is scaled by the 
nuclear overlap function \mbox{$\langle T_{\rm AA}\rangle$}  estimated through the Glauber model 
\cite{Miller:2007ri}.

ALICE (A Large Ion Collider Experiment) at the LHC is the experiment dedicated to study the physics of
nucleus-nucleus collisions. It is particularly well equipped to measure heavy-flavour hadrons
in different decay channels. At mid-rapidity, detectors in the central barrel, in a uniform magnetic
field of 0.5 T, provide high resolution track and vertex reconstruction, and precise particle identification 
for hadrons and for electrons. At forward-rapidity,
a muon spectrometer offers triggering and precise tracking of muons. A complete description of the
experiment can be found in \cite{ALICE}.
The heavy-flavour physics program of ALICE includes the following channels:
\begin{itemize}
\item semi-electronic decays studied via inclusive electron spectra measured at mid-rapidity;
\item semi-muonic decays at forward-rapidity;
\item full reconstruction of hadronic decay channels of charm mesons and baryons at mid-rapidity.
\end{itemize}

These will be described in the Sections \ref{hfe}, \ref{hfm}, and \ref{d2h}, respectively. Results from
both collision systems, proton-proton and Pb--Pb, at various center of mass energies will be presented
and discussed.

\section{Semi-leptonic decays of heavy-flavour hadrons at mid-rapidity}
\label{hfe}

One method to measure the production of heavy-flavour hadrons consists in
detecting the lepton produced in their semi-leptonic decays and analyzing
the inclusive lepton transverse momentum spectrum. Semi-leptonic decays offer
the advantage of a relatively large branching ratio, of the order of 10\% for both
charm and beauty hadrons.

\begin{figure}[tbh]
\begin{minipage}[b]{0.46\linewidth}
\centering
\includegraphics[width=0.97\textwidth]{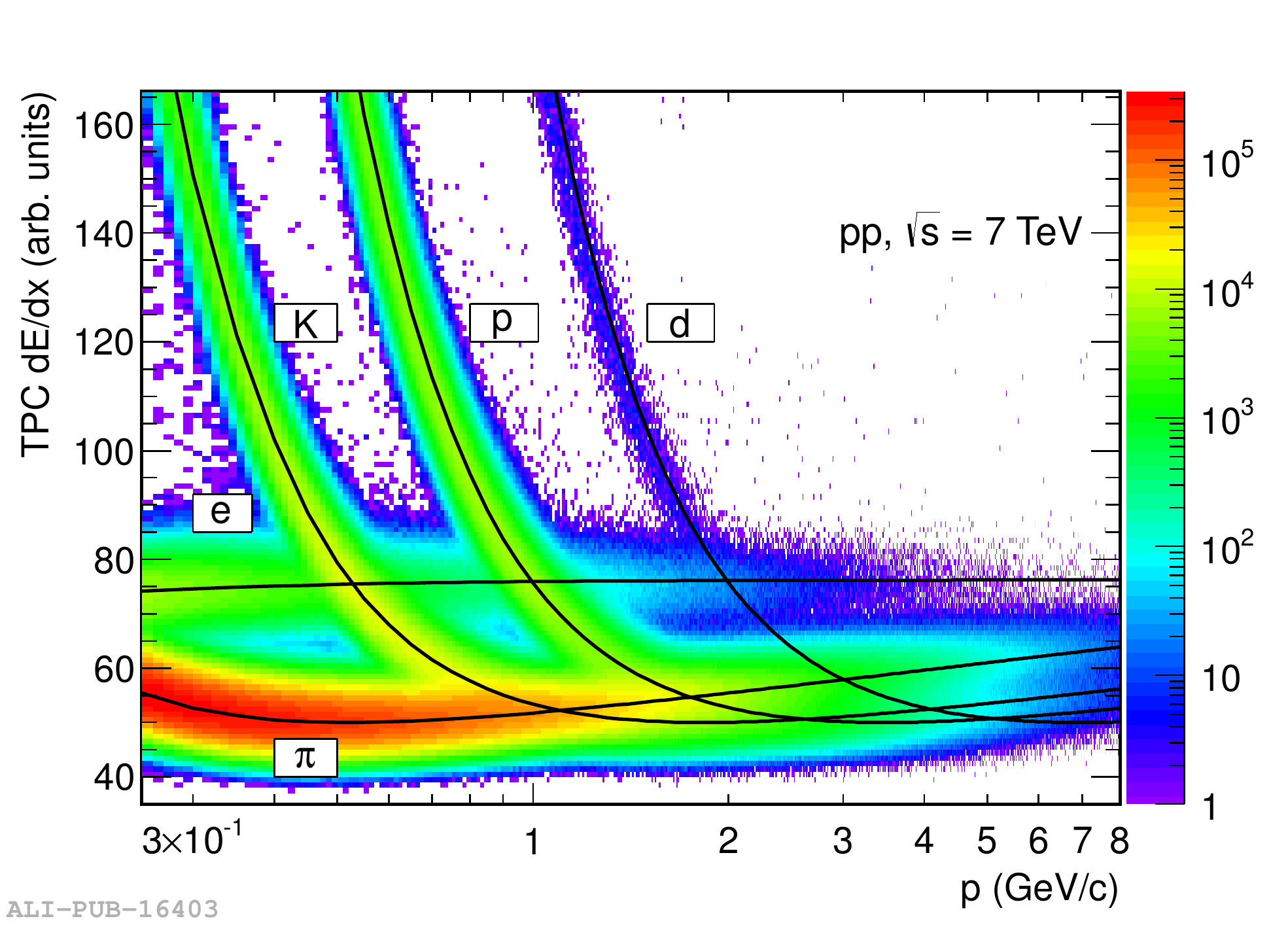}
\includegraphics[width=\textwidth]{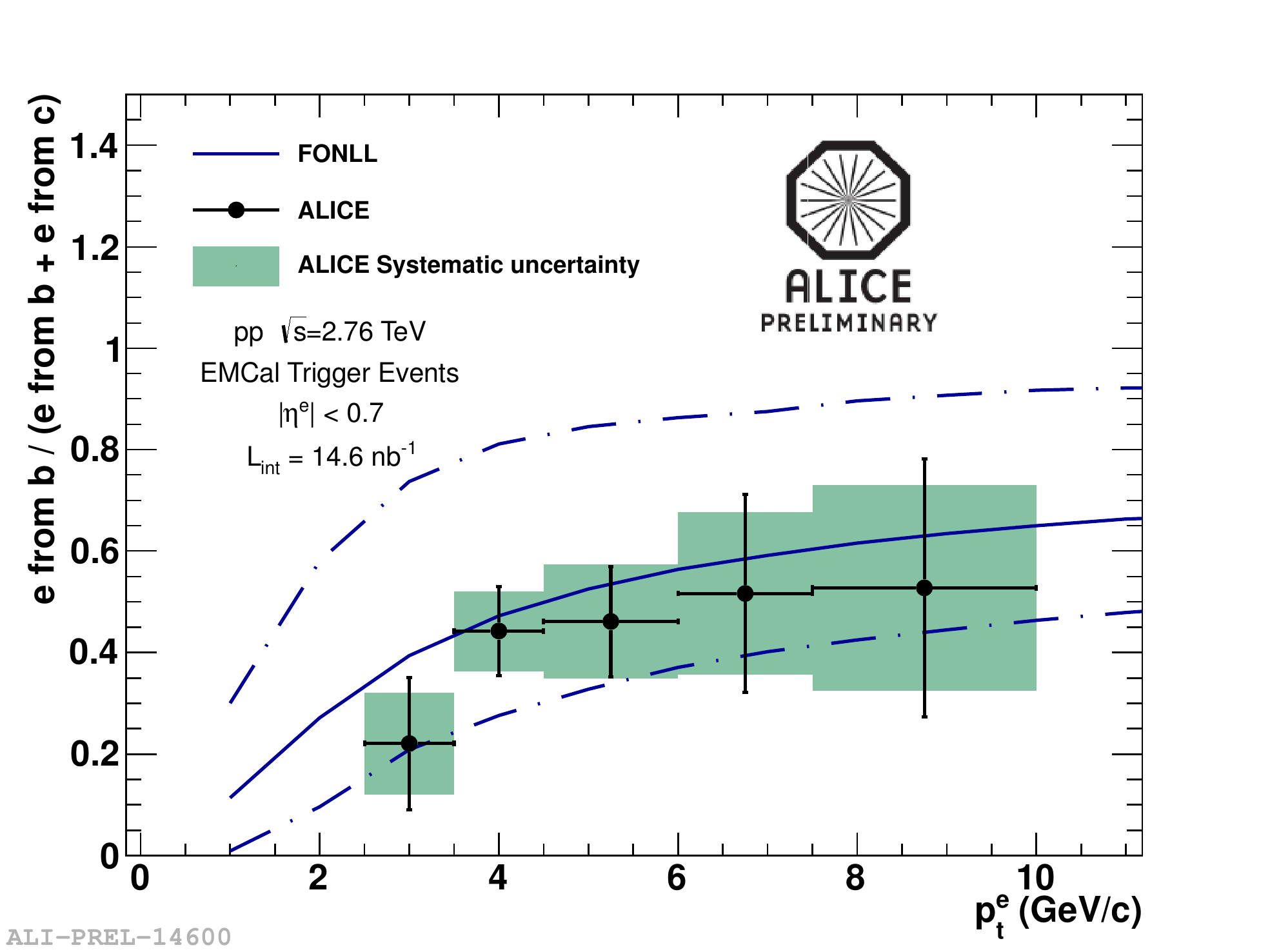}
\caption{(Top) Specific energy loss d$E$/d$x$
of charged tracks in the ALICE TPC (taken from \cite{HFEpaper7TeV}). 
Hadrons and electrons can be clearly identified.
(Bottom) Relative beauty contribution to the 
heavy-flavour electron yield, in \pp collisions at $\sqrt{s} = 2.76$~TeV,
compared with FONLL pQCD calculations.}
\label{fig:hfe2}
\end{minipage}
\hspace{0.4cm}
\begin{minipage}[b]{0.53\linewidth}
\centering
\includegraphics[width=\textwidth]{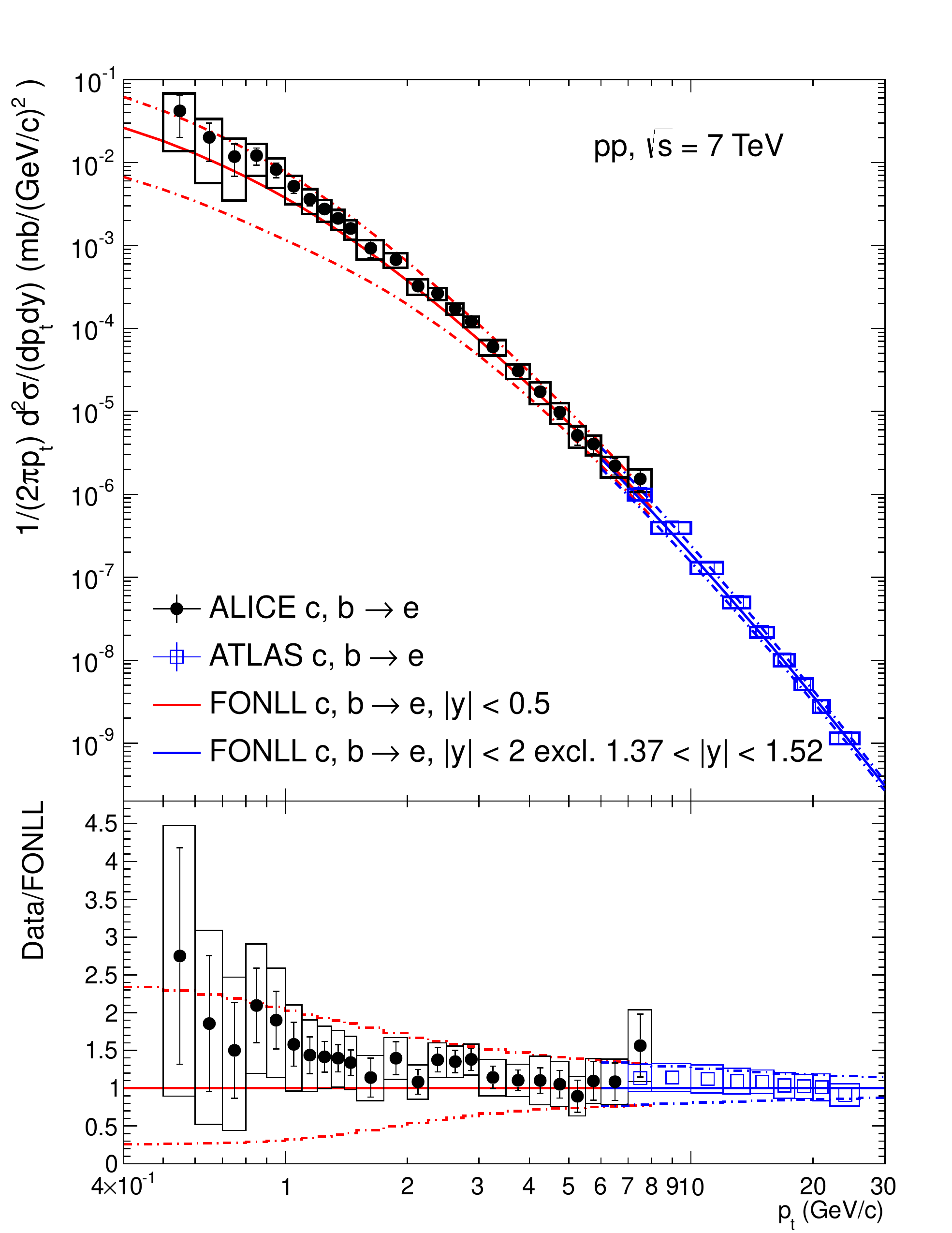}
\caption{Invariant differential production cross sections 
of electrons from heavy-flavour (charm and beauty) decays measured by ALICE~\cite{HFEpaper7TeV}, and 
ATLAS~\cite{Aad:2011rr} in \pp collisions at $\sqrt{s} = 7$~TeV. The measurements are 
compared to a
FONLL pQCD calculation.
\vspace*{2mm}}
\label{fig:hfe1}
\end{minipage}
\end{figure}

At mid-rapidity (extending from $\pm$0.5 to $\pm$0.8 depending on the detector
used in the measurement), electrons can be identified with high purity by a number
of Particle IDentification (PID) detectors. The central PID device used in all approaches
is the Time Projection Chamber (TPC), where the particle specific energy loss  d$E$/d$x$
is measured. The particle  d$E$/d$x$ as
a function of the track momentum is shown in Fig.~\ref{fig:hfe2} (top).
The ambiguities at low momentum, where the electron, kaon, proton and deuteron
lines cross, are resolved using information from the Time-of-Flight (TOF) detector.
At higher momenta, other detectors such as the Transition Radiation Detector (TRD)
or the ElectroMagnetic Calorimeter (EMCal) help suppressing the hadron contamination.

The inclusive electron spectrum contains contributions from many sources other than  
the heavy-flavour hadron decays. The most important are: Dalitz decays of light neutral  
mesons ($\pi^0$, $\eta$, $\omega$, $\eta$', $\phi$), photon conversions in 
the beam pipe and detector material, di-electron decays of vector mesons ($\rho$, 
$\omega$, $\phi$), heavy quarkonia, and direct radiation.
 
In a first approach, the spectrum composed by these background electrons is described  
by the so-called electron 
cocktail (see \cite{HFEpaper7TeV} for an extensive description of this method), which is  
statistically subtracted from the inclusive spectrum. This approach was used to obtain the 
production cross section of electrons from  
charm and beauty hadron decays in pp collisions at $\sqrt{s} = 7$~TeV, 
shown in Fig.~\ref{fig:hfe1}. Minimum bias events, for an integrated luminosity of 2.6~nb$^{-1}$,
were used.
The measurement is combined with a similar one by the ATLAS 
Collaboration~\cite{Aad:2011rr}, and compared to a pQCD calculation  
at fixed order with next-to-leading-log resummation (FONLL~\cite{fonll,fonll2,fonll3}). 
The theory prediction describes well the measurements over a very wide momentum range.
 
The combination with the ATLAS measurement helps to highlight one of the most important 
features of the ALICE apparatus, the capability to extend measurements to very low momenta, 
region in which a large part of the interesting physics resides, since the production cross section 
is largest. In fact, only by extending the measurement 
down to 0.5~GeV/$c$, can ALICE access $\approx$~50\% of the total charm cross section, and 
$\approx$~90\% of the beauty cross section, assuming FONLL transverse momentum shapes. 

This measurement treats inclusively the contributions from both charm and beauty hadrons.
The separation of the pure beauty component is achieved with two different analyses: one
is based on the azimuthal angular correlations of the electron with hadrons produced 
in its vicinity, and the second one on the larger separation of the electron from
the primary vertex for beauty hadron decays, because of their longer lifetime compared to
charm and other sources.

The former analysis exploints the fact that
the width of the near side correlation distribution is larger for beauty hadrons compared to
charm hadrons, due to different decay kinematics of the hadrons themselves \cite{Biritz:2009gm}.
By fitting the angular correlations measured in pp collisions at $\sqrt{s} = 2.76$~TeV,
the relative beauty contribution to the heavy-flavour hadron decay electron yield is determined. The result is 
shown in Fig.~\ref{fig:hfe2} (bottom). This is a first measurement of this type done in
pp collisions in ALICE. The data sample, corresponding to an  integrated luminosity of 14.6~nb$^{-1}$,
was chosen because an EMCal trigger
was used to select electrons above a cluster energy threshold of 3~GeV/$c$. 

Beauty hadrons have in average a longer lifetime (c$\tau \approx$ 500~$\mu$m) 
than charm hadrons and other background sources, therefore the
electrons from their semi-leptonic decays have a larger impact parameter with respect to the primary 
interaction vertex. The silicon vertex detector of ALICE, the Inner Tracking System (ITS), provides the
information on the track impact parameter with high spatial resolution, and an analysis based on the
electron separation from the primary interaction
is possible. This approach was used in the analysis of the pp data recorded at
$\sqrt{s} = 7$~TeV, to measure the $p_{\rm t}$-differential cross section of electrons from beauty hadron
decays only. First results were presented in \cite{Masciocchi:2011fu} and discussed in detail in 
\cite{Kweon:2012yn}.

Results from the electron analyses at mid-rapidity, at this conference, focused on measurements
in pp collisions. A first look at the nuclear modification factor of electrons from heavy-flavour
hadron decays was presented in \cite{Masciocchi:2011fu}. 

\section{Semi-leptonic decays of heavy-flavour hadrons at forward-rapidity}
\label{hfm}

An analogous measurement of heavy-flavour hadron semi-leptonic decays can be performed
at forward rapidity in ALICE, by reconstructing single muon tracks in the muon spectrometer. 
This consists of a passive front absorber, tracking chambers, and trigger chambers
located after an iron filter, and covers the pseudorapidity range -4$<y<$-2.5.

Muons are identified by requiring that a track reconstructed in the tracking chambers matches 
a corresponding track segment in the trigger chambers, to reject most of the reconstructed
hadrons, stopped in the iron wall. For the candidate muons,
the Distance-of-Closest-Approach (DCA) to the primary vertex is inspected, in order
to remove fake tracks and tracks from beam-gas interactions. The remaining background
contribution consists of muons from the decay in flight of light hadrons. 
In the case of proton-proton collisions, the contribution from light hadron decays can be
safely estimated with Monte Carlo simulations, and it is then subtracted from the measured
inclusive spectrum. 

\begin{figure}[tbh]
\centering
\includegraphics[width=0.4\textwidth]{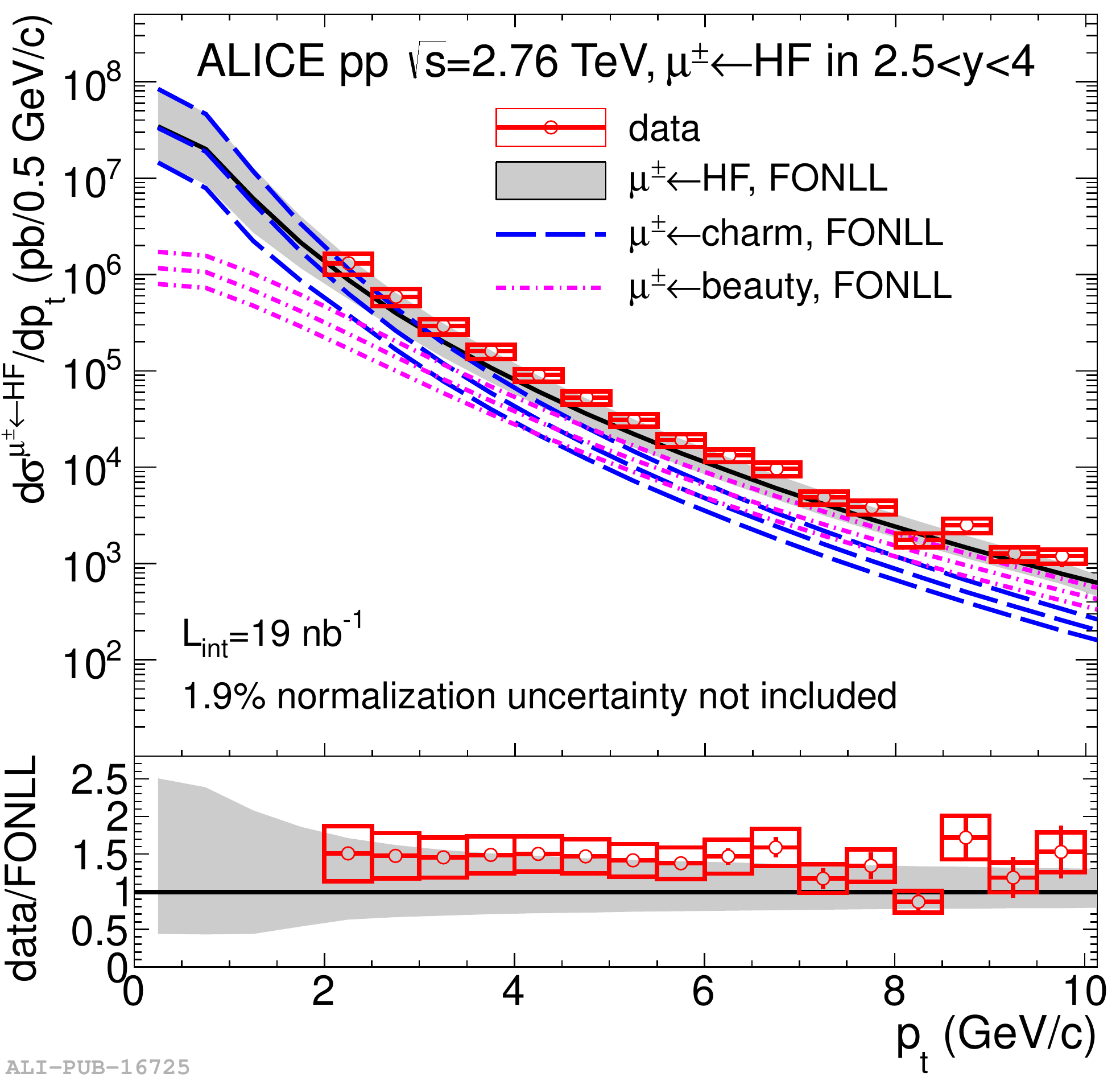}
\caption{Production cross section of muons from heavy-flavour hadron decays measured in 
proton-proton collisions at $\sqrt{s}$ = 2.76~TeV \cite{Stocco:2012ci,Abelev:2012qh}. 
The measurement is compared to a
FONLL prediction \cite{fonll,fonll2}. The ratio data to FONLL is shown in the
lower panel.}
\label{fig:hfm1}
\end{figure}

This analysis strategy was applied to reconstruct the $p_{\rm t}$-differential
production cross section of muons from heavy-flavour hadron decays in pp collisions both at
$\sqrt{s}$ = 7~TeV \cite{Abelev:2012pi}, and at 2.76~TeV. The latter result was newly 
presented at this conference and reported in detail in these proceedings \cite{Stocco:2012ci}
and in the publication \cite{Abelev:2012qh}. The resulting cross
section is compared to a FONLL prediction for the summed contribution from charm and
beauty hadron decays, and is shown in Fig.~\ref{fig:hfm1}. The theoretical prediction describes well the data.

A simple Monte Carlo approach is however not sufficient for Pb--Pb collisions, where medium effects
are expected to affect the light hadron spectra. A data driven method is used, which takes pion and kaon
distributions measured in the ALICE central barrel and extrapolates them to forward rapidities. 
A $\pm$100\% systematic uncertainty is assigned in this step, because of the unknown rapidity
dependence of the medium effects on the light hadrons.
The extrapolated spectra
are then used to generate the decay muons in a full simulation of the detector setup. The systematic
uncertainty on the background contribution which is subtracted allows for the application of this method
for transverse momentum values above 4~\gevc (2~\gevc in pp collisions).

\begin{figure}[tbh]
\centering
\includegraphics[width=0.7\textwidth]{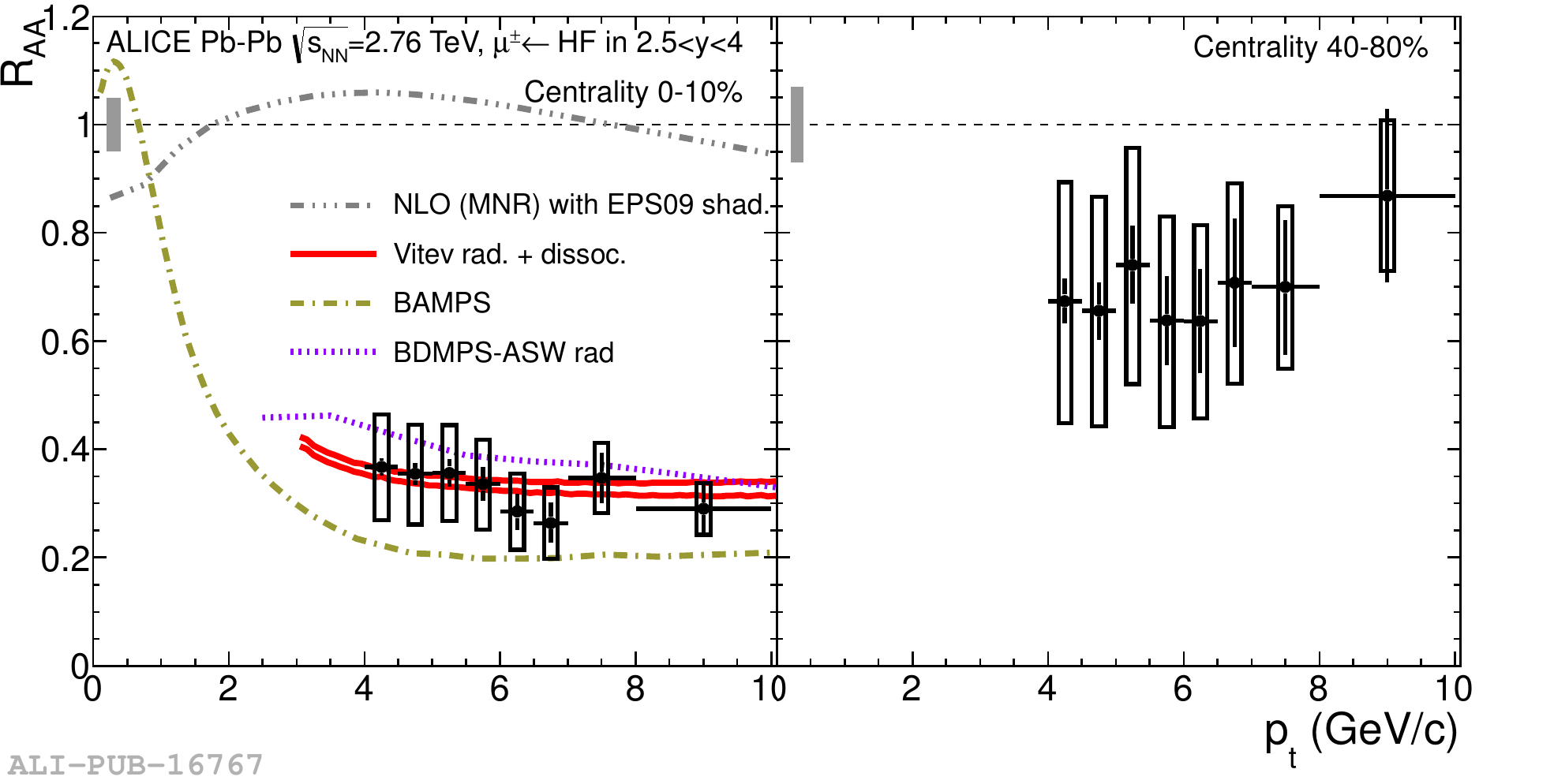}
\caption{Nuclear modification factor as a function of the transverse momentum
of the muons from heavy-flavour hadron decays, in the 0-10\% most central collisions (left) and
in the 40-80\% peripheral collisions (right)  \cite{Stocco:2012ci,Abelev:2012qh}. 
The models compared to the data are discussed in the text.}
\label{fig:hfm2}
\end{figure}

Pb--Pb collisions at {\mbox {$\sqrt{s_{\rm NN}}$ = 2.76 TeV}} recorded in Fall 2010 were
analyzed and the data driven method was used to obtain a \pt spectrum of muons from heavy-flavour
hadron decays for different centrality intervals of the ion-ion collisions. 16.6 million minimum bias
trigger events were analyzed, for an integrated luminosity of 2.7~$\mu$b$^{-1}$.
The spectra were used
to calculate the nuclear modification factor $R_{\rm AA}$, defined in equation \ref{eq:raa}. The result is 
shown in Fig.\ref{fig:hfm2}, on the left for the 10\% most central collisions, and on the right for 
the 40-80\% peripheral collisions. A strong suppression (by a factor 3-4) of the 
muons from
heavy-flavour hadron decays
can be seen in the most central collisions, and is independent on the muon transverse
momentum in the measured range. The suppression is much reduced in the peripheral collisions
where lower energy densities are reached.

The nuclear modification factor measured in the most central collisions (Fig.~\ref{fig:hfm2}, left)
is compared to model predictions. First of all the possible effect of nuclear shadowing is estimated
by using perturbative calculations by Mangano,
Nason and Ridolfi \cite{Mangano:1992kq}   and the EPS09NLO \cite{Eskola:2009uj}   
parameterization of the shadowing. This initial state 
effect reduces the parton distribution functions for partons carrying a fraction of
the nucleon momentum smaller than 10$^{-2}$. The calculation is represented by the grey dotted-dashed
curve on top, labelled NLO: the shadowing effect is small and the comparison to the data suggests
the strong suppression to be dominantly a final state effect.

The other curves shown in the same panel represent the nuclear modification factor calculated with
models which implement collisional (BAMPS) \cite{Uphoff:2012gb}, radiative (BDMPS-APW) 
\cite{Armesto:2005iq}, and radiative with in-medium hadron formation and dissociation
 \cite{Sharma:2009hn} energy loss.
Good agreement is found in the two cases where
radiative energy loss is included, while BAMPS underestimates the muon $R_{\rm AA}$.
For a more detailed description and discussion of
the  muon results, see ~\cite{Abelev:2012pi,Stocco:2012ci,Abelev:2012qh}.

\section{Hadronic decays of charm hadrons}
\label{d2h}

Hadrons containing a charm quark can be selected by fully reconstructing their decays 
with only hadrons in the final state.
Charged tracks are reconstructed and identified at mid-rapidity in the
ITS, the TPC and the TOF detectors, and combined into candidate charm particles
with an invariant mass analysis. The possibility to detect charm signals via exclusive 
reconstruction of hadronic decay channels
with very large statistics was largely improved at the LHC by
the much increased
heavy-flavour production cross sections compared to collisions at lower energy, and thanks
to the new generation of silicon vertex detectors. ALICE is especially well equipped for these
measurements because, in addition to the high resolution track and vertex reconstruction,
it also profits from very good particle identification over a large momentum range.
The combinatorial background rejection is largely improved by the use of the 
information on the particle identification.
The issue is particularly important in the high
multiplicity environment of central Pb--Pb collisions, where a variety of decay channels
could be reconstructed by ALICE.

\subsection{Charm-hadron production cross section in pp}
\label{d2h_pp}
 
The main decay channels studied in ALICE are: D$^0 ~\rightarrow$~K$^-~\pi^+$, 
D$^+~\rightarrow$~K$^-~\pi^+~\pi^+$, D*$^{+}~\rightarrow$~D$^0 ~\pi^+$. Via
these decay channels, the $p_{\rm t}$-differential production cross section for prompt charm mesons (where
prompt indicates that they are produced in the primary interaction and do not come
from the decay of a beauty hadron) has been measured in pp collisions at $\sqrt{s}$ = 7~TeV 
and at 2.76~TeV, as reported in \cite{ALICE:2011aa} and \cite{aliced:2012sx}, respectively.
As in the case of the semi-leptonic decay measurements, the resulting production cross
sections are well described by pQCD predictions, both from FONLL~\cite{fonll,fonll2,fonll3}
and from a massive variable flavor number scheme (GM-VFNS \cite{Kniehl:2005st,Kniehl:2005mk}).

The $p_{\rm t}$-differential  production cross sections were extrapolated to the full phase
space to determine the total c$\bar{\rm c}$ production cross section at the two collision
energies \cite{aliced:2012sx}.

Other charm hadrons are investigated in pp collisions. A measurement of the D$_{\rm s}$ production 
cross section, in the decay channel D$_{\rm s} ~\rightarrow$~K$^+$~K$^-~\pi^+$,
was presented
\cite{Geuna:2012tu,aliceds:2012ik}. A signal is observed for the $\Lambda_{\rm c}$ baryon in two decay channels:
p~K$^-~\pi^+$ and pK$_{\rm s}^0$. 

The variety of measured charm hadrons and decay channels provide a stringent cross-check
of all heavy-flavour measurements. The investigation of the $\Lambda_{\rm c}$ baryon
is very interesting in view of future possibilities to study in heavy-ion collisions the meson to
baryon ratio also in the charm sector.

\subsection{Nuclear modification factor of charm hadrons}
\label{d2h_raa}
\begin{figure}[tbh]
\centering
\includegraphics[width=0.87\textwidth]{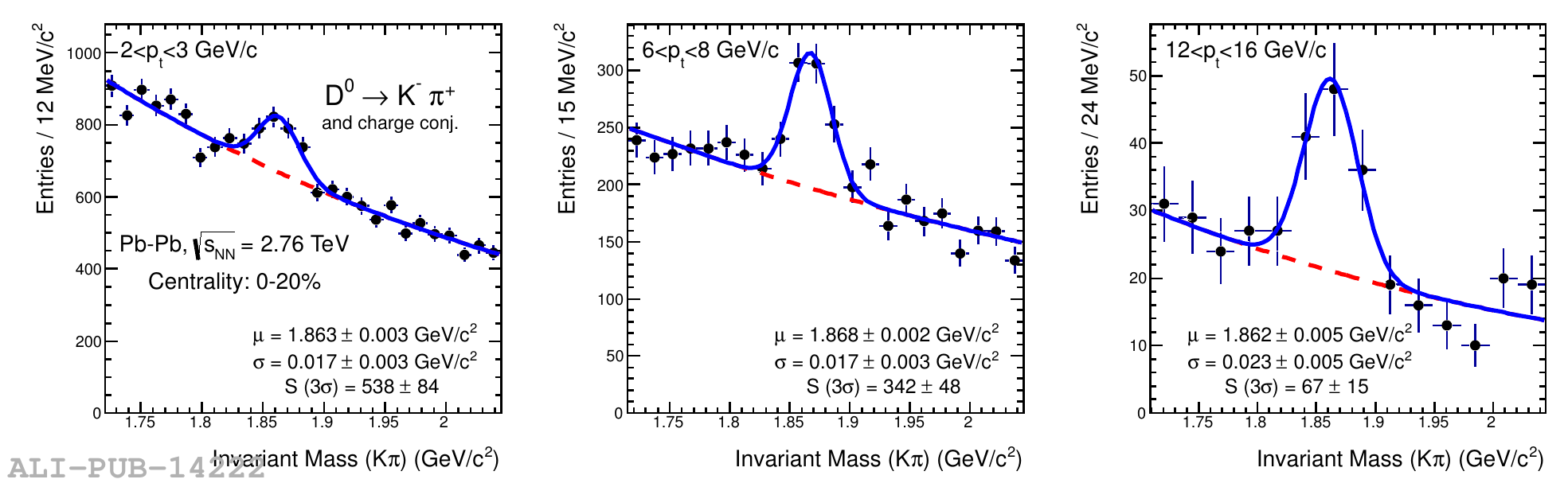}
\caption{Invariant mass distributions for D$^0$ candidates
 in three \pt intervals for the 0--20\%
most central Pb--Pb collisions \cite{ALICE:2012ab}.}
\label{fig:d2hmass}
\end{figure}

The reconstruction of the hadronic decays of charm hadrons in Pb--Pb collisions at 
{\mbox {$\sqrt{s_{\rm NN}}$ = 2.76 TeV}} is a challenging task, because of the high combinatorial
background, due to the very high track multiplicities. Figure~\ref{fig:d2hmass} shows the
invariant mass distributions for the decay D$^0 ~\rightarrow$~K$^-~\pi^+$ in three different \pt bins.

Also in Pb--Pb collisions the three main D meson decay channels discussed in \ref{d2h_pp} are
reconstructed, and differential \pt spectra are measured over the range 2$ < $\pt$ < $16~\gevc.
Results from the 2010 sample of Pb--Pb minimum bias events, for an integrated luminosity of 
about 2.1 $\mu$b$^{-1}$, are presented.
Prompt D meson yields are obtained by subtracting the contribution of D mesons from B
decays. This is evaluated using the FONLL estimate of the beauty production cross section,
and the B$\rightarrow$D decay kinematics. The contribution is then renormalized by the
average nuclear overlap function in each centrality interval, and the nuclear modification factor
of the D mesons from B decays, assuming it equal to that of prompt D mesons. All details of
the analysis in Pb--Pb can be found in \cite{ALICE:2012ab}.

\begin{figure}[tbh]
\centering
\includegraphics[width=0.7\textwidth]{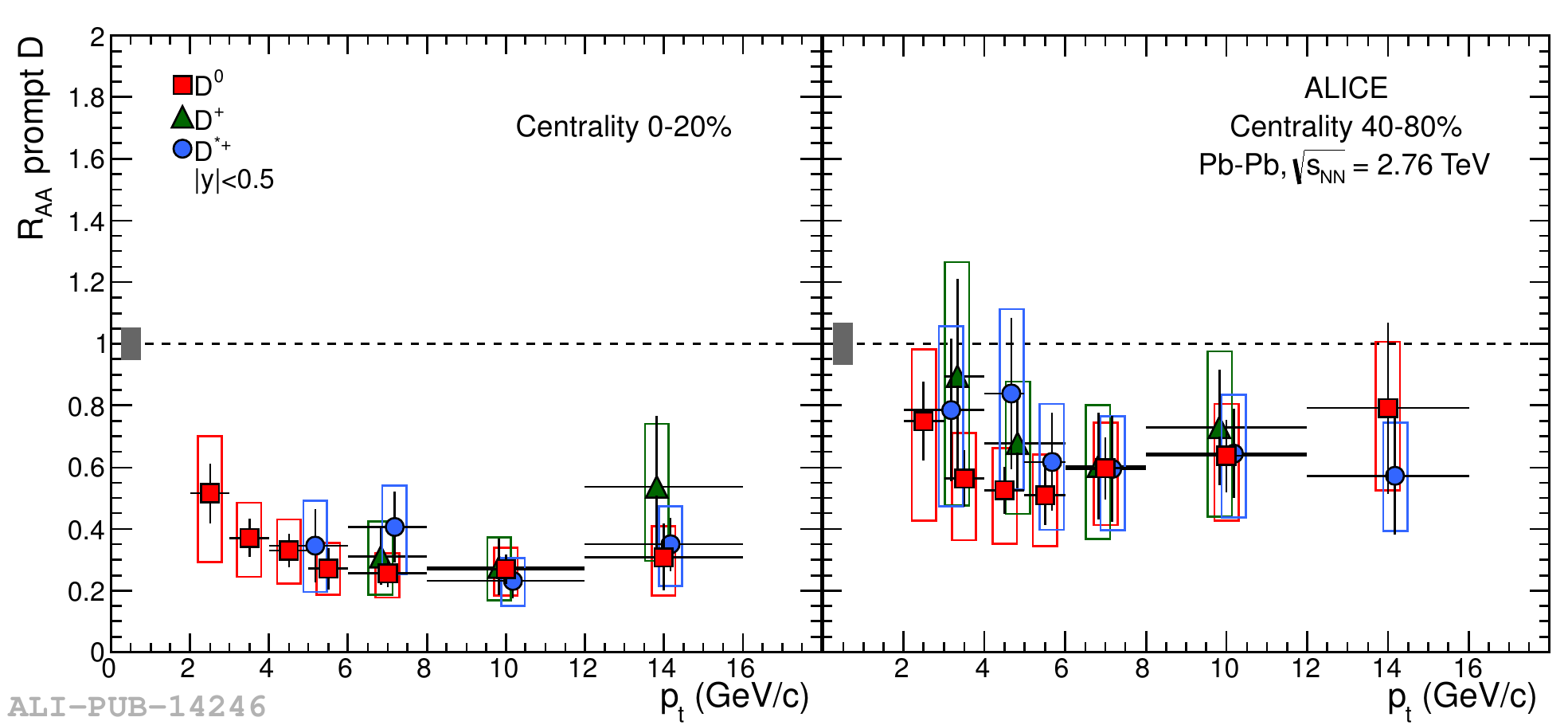}
\caption{Nuclear modification factor for prompt D$^0$, D$^+$, and D*$^+$ as a function
of the meson transverse momentum, in the 0-20\% (left) and 40-80\% (right) centrality
intervals \cite{ALICE:2012ab}.}
\label{fig:d2hraapt}
\end{figure}

The prompt D$^0$, D$^+$, and D*$^+$ $p_{\rm t}$-differential yields in the 0-20\% and 40-80\%
centrality classes are then used to compute the nuclear modification factor, shown in 
Fig.~\ref{fig:d2hraapt} as a function of the D meson transverse momentum.

The results from the three D mesons are compatible, and show a suppression by a 
factor of 3-4 in the most central events, with no significant dependence on the transverse
momentum above 5~\gevc. The suppression factor increases going from more peripheral
to more central collisions. 

\begin{figure}[tbh]
\centering
\includegraphics[width=0.37\textwidth]{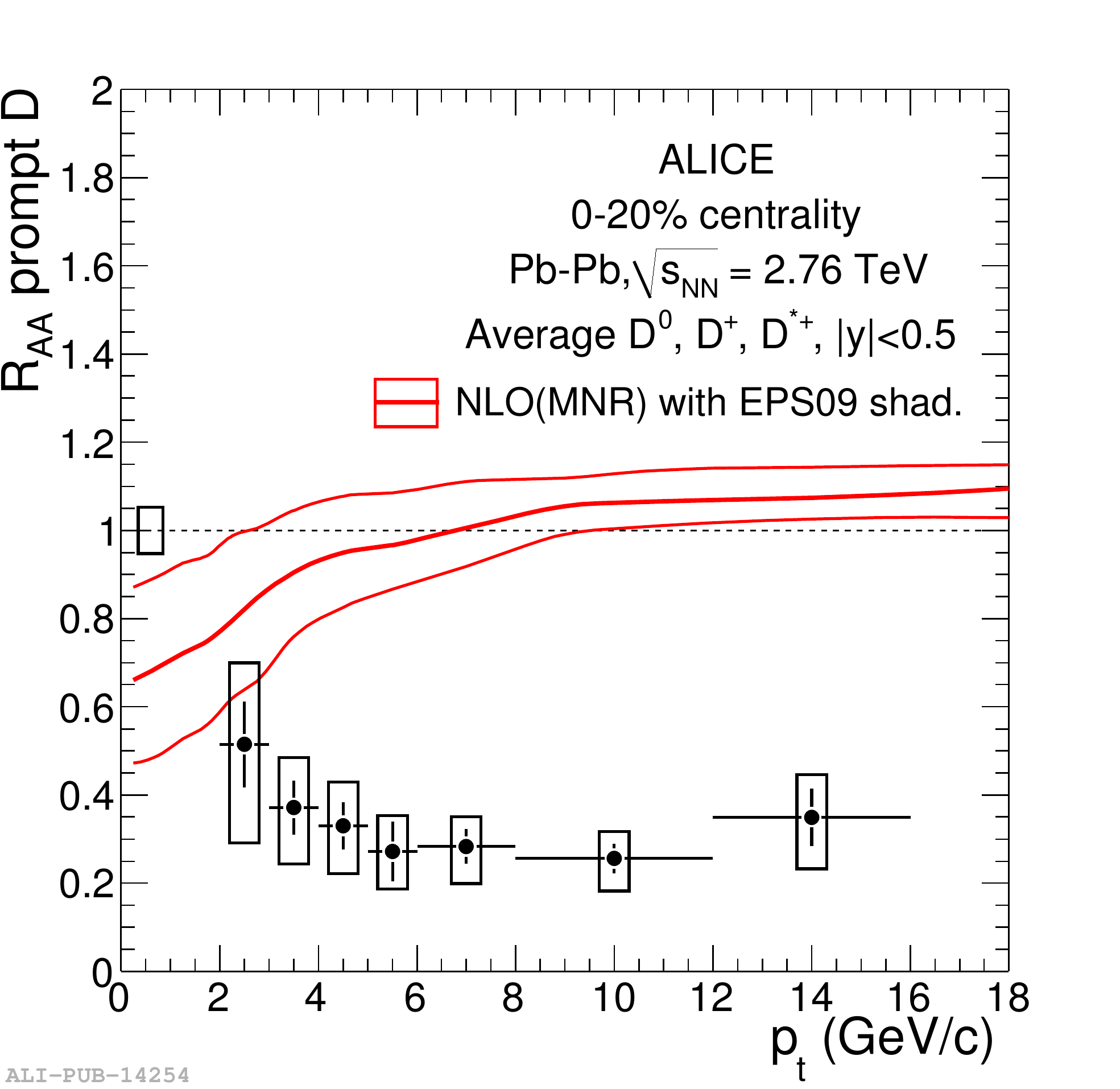}
\hspace{0.5cm}
\includegraphics[width=0.39\textwidth]{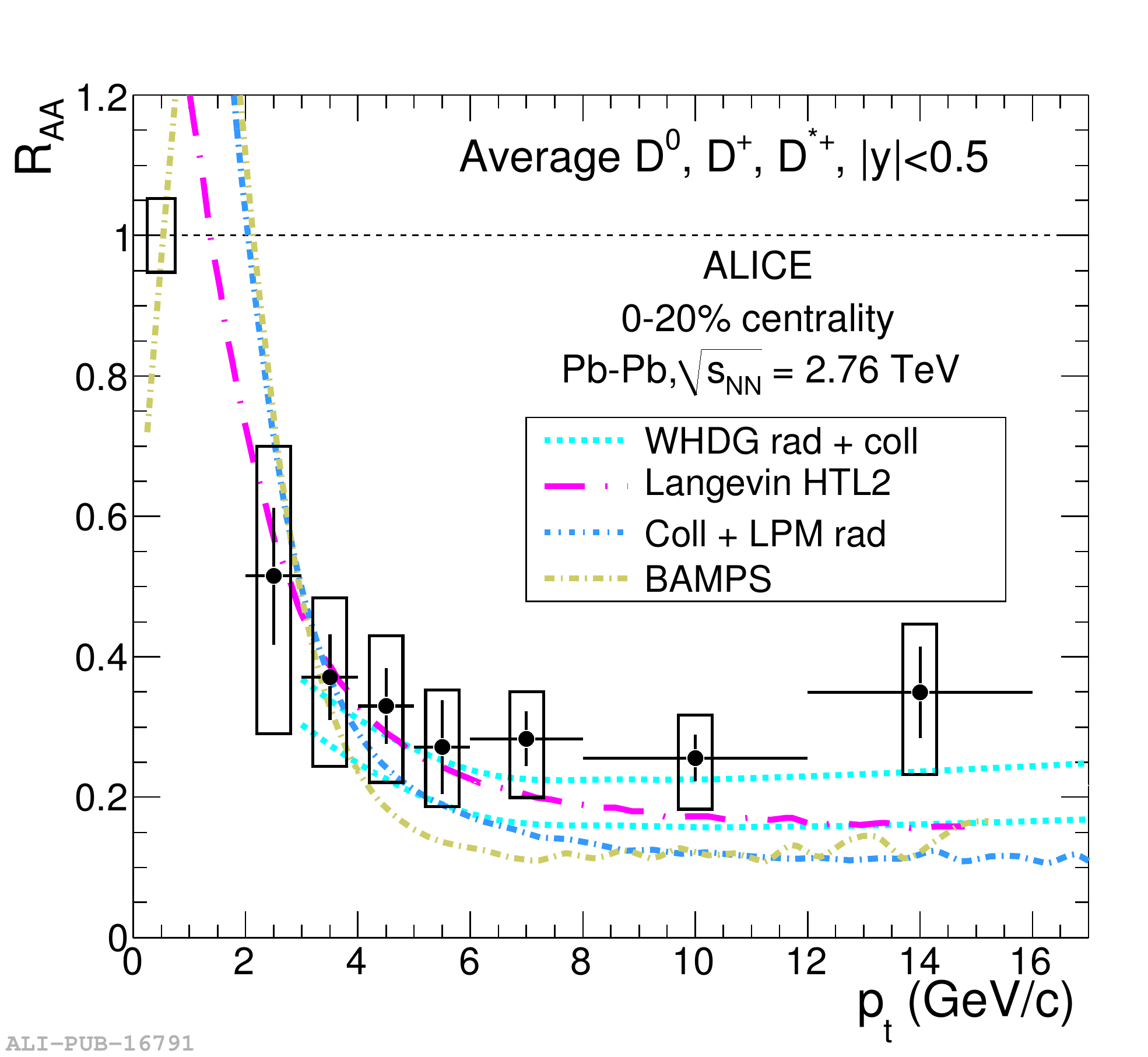}
\caption{The average D meson $R_{\rm AA}$ is compared with a calculation of nuclear shadowing
(left) \cite{ALICE:2012ab} and various parton energy loss models: WHDG rad + coll \cite{Horowitz:2011cv},
Langevin HTL2 \cite{Alberico:2011zy}, Coll + LPM rad \cite{Gossiaux:2009mk,Gossiaux:2010yx}, 
and BAMPS \cite{Uphoff:2012gb}.}
\label{fig:d2hmodel}
\end{figure}

The averaged D meson $R_{\rm AA}$ is compared with a calculation including the effects
of nuclear shadowing, based on perturbative calculations by Mangano,
Nason and Ridolfi \cite{Mangano:1992kq}   and the EPS09NLO \cite{Eskola:2009uj}   
parameterization, similarly to the muon case discussed in the previous section. The
comparison is shown in Fig.~\ref{fig:d2hmodel} (left): above 6~GeV/$c$, the shadowing-induced
effect on $R_{\rm AA}$ is small and cannot explain the measurement.
This suggests once again that the strong suppression of heavy flavours observed in data is a final state effect.

The parton energy loss in the medium is described by a number of different models, which 
calculate the charm nuclear modification factor. Some of the predictions are shown in Fig.~\ref{fig:d2hmodel}
(right). As commented in the muon case, models including radiative components of
the energy loss describe reasonably well the measured suppression.

\subsection{Elliptic flow of D mesons}
\label{d2h_flow}
In addition to the parton energy loss in the QGP, another very important question about heavy
quarks concerns their degree of thermalization inside the medium and the participation to its
collective motion. A study of the azimuthal anisotropy of the prompt D meson production 
brings information, at low transverse momentum, on the charm thermalization and, at high
momentum, on the path-length dependence of the parton energy loss. 
The dominant component of the anisotropic distribution is called “elliptic flow”
and is commonly quantified by the second coefficient, $v_2$,
of a Fourier decomposition of the azimuthal distribution
of observed particles relative to the event-plane angle.
The flow measurement
was made possible by the larger statistics recorded in the 2011 Pb--Pb data taking period.
A centrality trigger enhanced the statistics in the 30-50\% centrality interval to
almost 10$^7$ events.

The event plane is
determined from the distribution of charged tracks in the TPC. 
The elliptic flow of D$^0$ mesons was calculated by comparing the signal found in two complementary
angular regions of the transversal plane, one called ``in plane'', the other ``out of plane''. The invariant
mass analysis provides the number of signal candidates found in the two regions ($N_{IN}, N_{OUT}$)
and the elliptic flow can be obtained as:
$v_2~=~\frac{\pi}{4} \frac{N_{IN}-N_{OUT}}{N_{IN}+N_{OUT}} \frac{1}{r}$, where $r$ is the event plane resolution.
Details of this analysis are discussed in \cite{Ortona:2012gx}.

\begin{figure}[tbh]
\centering
\includegraphics[width=0.4\textwidth]{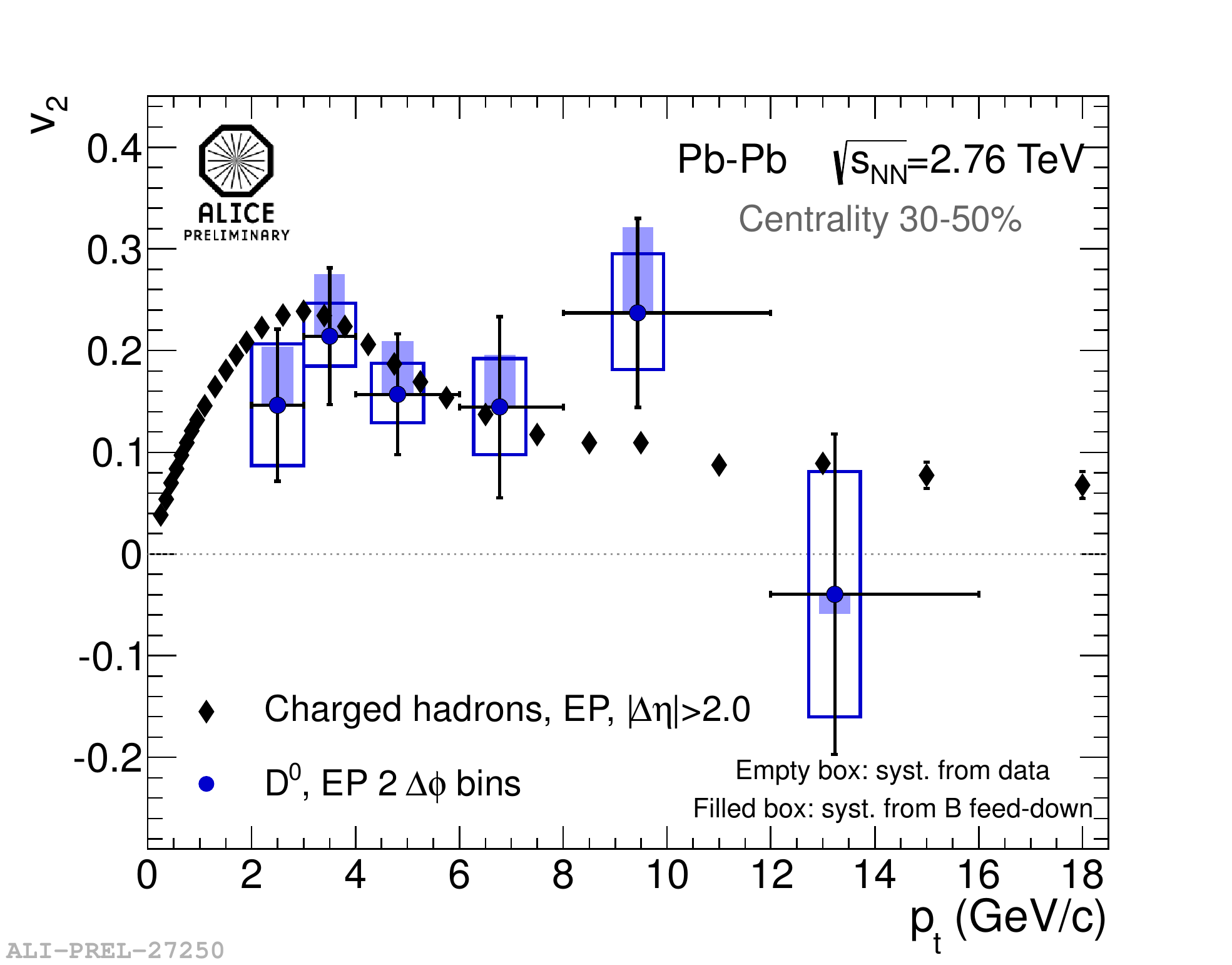}
\includegraphics[width=0.4\textwidth]{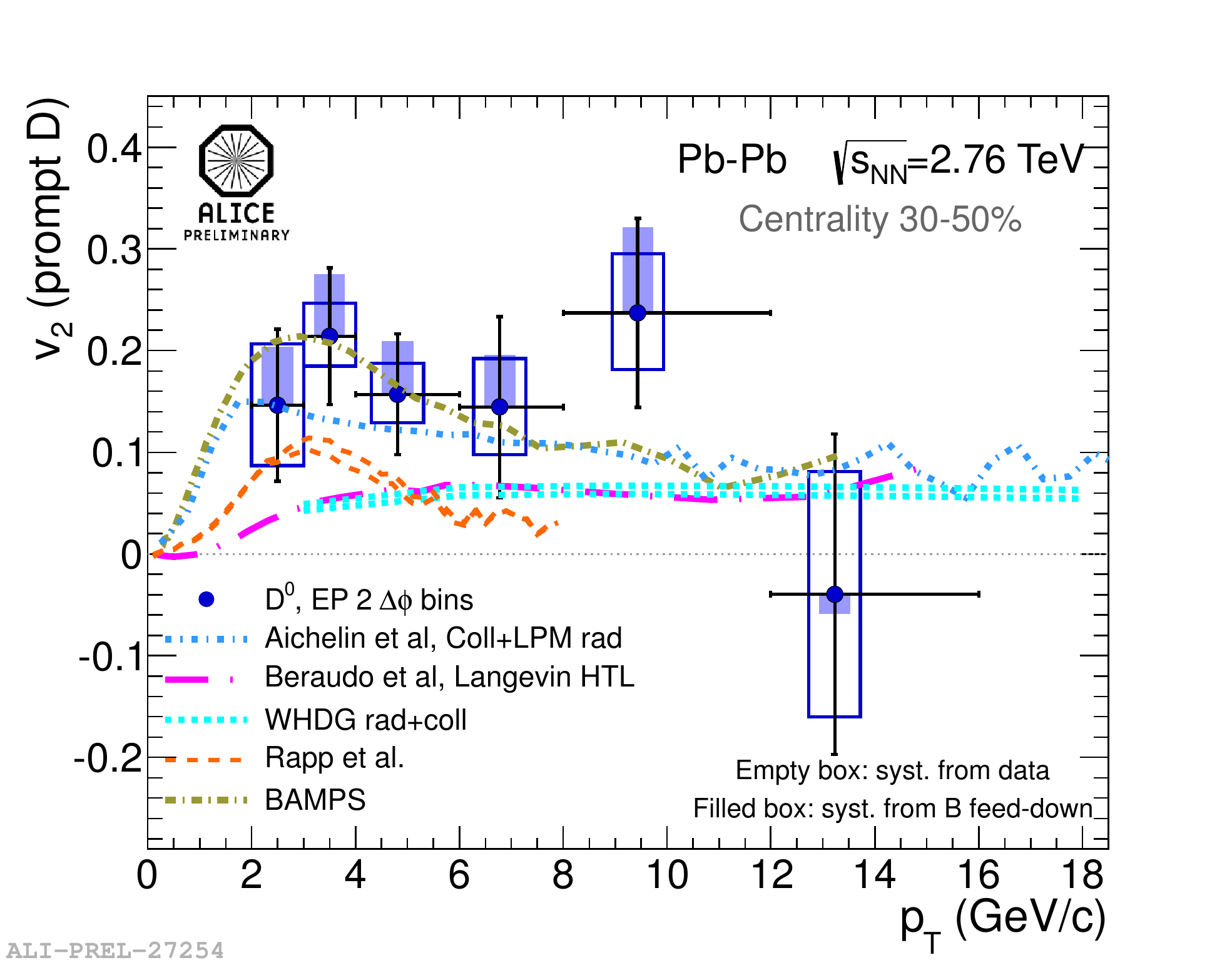}
\caption{(Left) D$^0$ elliptic flow measured in the 30-50\% centrality class \cite{Ortona:2012gx}, 
compared to the elliptic flow of charged particles measured in ALICE \cite{Aamodt:2010pa}.
(Right) Comparison to the same models for which a description of the D meson $R_{\rm AA}$
was given in Fig.~\ref{fig:d2hmodel}.}
\label{fig:flow}
\end{figure}

The resulting $v_2$ of D$^0$ mesons as a function of the transverse momentum is shown in Fig.~\ref{fig:flow} (left).
A non-zero elliptic flow (with a 3$\sigma$ significance) is found in the range 2$<$\pt$<$6~\gevc.
The result is of the same magnitude of the $v_2$ of charged tracks measured in ALICE, within the 
large uncertainties.
The D$^0$ result is confirmed by a similar measurement with the D$^+$ in the same centrality bin,
and by different methods to compute $v_2$.  The figure on the right side shows a comparison of 
the measured elliptic flow with predictions by the same models already used in Fig.~\ref{fig:d2hmodel}
in comparison to the D meson $R_{\rm AA}$. With the availability of the many results on nuclear modification
factors and elliptic flow of heavy flavours, it is
now very important that theoretical models
provide a simultaneous and coherent description of the transport and
energy loss of the
heavy quarks in the medium.

\section{Summary}
\label{conclusion}
After the first two years of data taking with proton-proton and Pb--Pb collisions at the LHC, at the highest ever reached energies,
ALICE has presented a very rich collection of physics results on the production of heavy
quarks in hadronic collisions, and their modification in the presence of the strongly-interacting, deconfined
medium produced in heavy-ion collisions. Production cross sections measured in pp both for charm and for beauty
are very well described by pQCD predictions, and provide the reference for measurements in Pb--Pb. In the most
central heavy-ion collisions, a strong suppression by a factor 3-4 is measured for heavy-quark hadrons, in the
intermediate \pt region. The suppression exhibits a clear
centrality dependence and no strong \pt dependence in the range 5$<p_{\rm t}<15$~\gevc.
A non-zero elliptic flow of D mesons is measured at low transverse momentum, in semi-central collisions.
One of the main goals for the near future is to extend the \pt range of the measurements, both
to higher momenta, and to very low \pt where very interesting physics is to be understood. At low momentum,
in fact, the heavy-flavour production cross section is largest and, in heavy-ion collisions, the quenched 
heavy-flavour hadrons have to accumulate. Furthermore, it will soon be possible 
to  isolate and investigate the beauty component also in Pb--Pb collisions.








\bibliographystyle{unsrt}
\bibliography{sma_HP2012_aliceHF}

\begin{thebibliography}{10}

\bibitem{Miller:2007ri}
M.~L. Miller, K.~Reygers, S.~J. Sanders, and P.~Steinberg.
\newblock {\em Ann.Rev.Nucl.Part.Sci.}, 57:205--243, 2007.

\bibitem{ALICE}
K.~Aamodt et~al.
\newblock {\em JINST}, 3:S08002, 2008.

\bibitem{HFEpaper7TeV}
B.~Abelev et~al.
\newblock arXiv1205.5423 [hep-ex].

\bibitem{Aad:2011rr}
G.~Aad et~al.
\newblock {\em Phys. Lett.}, B707:438--458, 2012.

\bibitem{fonll}
M.~Cacciari, M.~Greco, and P.~Nason.
\newblock {\em JHEP}, 9805:007, 1998.

\bibitem{fonll2}
M.~Cacciari, S.~Frixione, and P.~Nason.
\newblock {\em JHEP}, 0103:006, 2001.

\bibitem{fonll3}
M.~Cacciari et~al.
\newblock arXiv:1205.6344 [hep-ph].

\bibitem{Biritz:2009gm}
B.~Biritz.
\newblock {\em Nucl.Phys.}, A830:849C--852C, 2009.

\bibitem{Masciocchi:2011fu}
S.~Masciocchi for~the ALICE~Coll.
\newblock {\em J.Phys.}, G38:124069, 2011.

\bibitem{Kweon:2012yn}
M.~Kweon for~the ALICE~Coll.
\newblock These proceedings. arXiv:1208.5411 [nucl-ex].

\bibitem{Abelev:2012pi}
B.~Abelev et~al.
\newblock {\em Phys.Lett.}, B708:265--275, 2012.

\bibitem{Stocco:2012ci}
D.~Stocco for~the ALICE~Coll.
\newblock These proceedings. arXiv:1208.6171 [nucl-ex].

\bibitem{Abelev:2012qh}
B.~Abelev et~al.
\newblock arXiv:1205.6443 [nucl-ex].

\bibitem{Mangano:1992kq}
M.~L. Mangano, P.~Nason, and G.~Ridolfi.
\newblock {\em Nucl.Phys.}, B405:507--535, 1993.

\bibitem{Eskola:2009uj}
K.J. Eskola, H.~Paukkunen, and C.A. Salgado.
\newblock {\em JHEP}, 0904:065, 2009.

\bibitem{Uphoff:2012gb}
J.~Uphoff, O.~Fochler, Z.~Xu, and C.~Greiner.
\newblock These proceedings. arXiv:1205.4945 [nucl-ex].

\bibitem{Armesto:2005iq}
N.~Armesto, A.~Dainese, C.~A. Salgado, and U.~A. Wiedemann.
\newblock {\em Phys.Rev.}, D71:054027, 2005.

\bibitem{Sharma:2009hn}
R.~Sharma, I.~Vitev, and B.~Zhang.
\newblock {\em Phys.Rev.}, C80:054902, 2009.

\bibitem{ALICE:2011aa}
B.~Abelev et~al.
\newblock {\em JHEP}, 1201:128, 2012.

\bibitem{aliced:2012sx}
B.~Abelev et~al.
\newblock {\em JHEP}, 1207:191, 2012.

\bibitem{Kniehl:2005st}
B.A. Kniehl, G.~Kramer, I.~Schienbein, and H.~Spiesberger.
\newblock {\em AIP Conf.Proc.}, 792:867--870, 2005.

\bibitem{Kniehl:2005mk}
B.A. Kniehl, G.~Kramer, I.~Schienbein, and H.~Spiesberger.
\newblock {\em Eur.Phys.J.}, C41:199--212, 2005.

\bibitem{Geuna:2012tu}
C.~Geuna for~the ALICE~Coll.
\newblock These proceedings. arXiv:1209.0382 [hep-ex].

\bibitem{aliceds:2012ik}
B.~Abelev et~al.
\newblock arXiv:1208.1948 [hep-ex].

\bibitem{ALICE:2012ab}
B.~Abelev et~al.
\newblock arXiv:1203.2160 [hep-ex].

\bibitem{Horowitz:2011cv}
W.A. Horowitz and M.~Gyulassy.
\newblock {\em J.Phys.}, G38:124114, 2011.

\bibitem{Alberico:2011zy}
W.M. Alberico, A.~Beraudo, A.~De~Pace, A.~Molinari, M.~Monteno, et~al.
\newblock {\em Eur.Phys.J.}, C71:1666, 2011.

\bibitem{Gossiaux:2009mk}
P.B. Gossiaux, R.~Bierkandt, and J.~Aichelin.
\newblock {\em Phys.Rev.}, C79:044906, 2009.

\bibitem{Gossiaux:2010yx}
P.B. Gossiaux, J.~Aichelin, T.~Gousset, and V.~Guiho.
\newblock {\em J.Phys.}, G37:094019, 2010.

\bibitem{Ortona:2012gx}
G.~Ortona for~the ALICE~Coll.
\newblock These proceedings. arXiv:1207.7239 [hep-ex].

\bibitem{Aamodt:2010pa}
K.~Aamodt et~al.
\newblock {\em Phys.Rev.Lett.}, 105:252302, 2010.

\end{thebibliography}





\end{document}